\documentstyle[aps,12pt]{revtex}
\newcommand{\bn}{\mbox{\boldmath $n$\unboldmath}}
\newcommand{\bl}{\mbox{\boldmath $l$\unboldmath}}
\newcommand{\bm}{\mbox{\boldmath $m$\unboldmath}}
\newcommand{\cD}{\cal D}
\newcommand{\cL}{\cal L}
\newcommand{\cLd}{{\cal L}^{\dagger}}
\newcommand{\tkr}{2\kappa (r-r_H)}
\newcommand{\sta}{\sin\theta}
\newcommand{\cta}{\cos\theta}

\newcommand{\coa}{\cot\theta}
\newcommand{\sqd}{\sqrt{2}}

\newcommand{\pr}{\frac{\partial}{\partial r}}
\newcommand{\pv}{\frac{\partial}{\partial v}}
\newcommand{\pta}{\frac{\partial}{\partial \theta}}
\newcommand{\pvi}{\frac{\partial}{\partial \varphi}}

\newcommand{\spr}{\frac{\partial}{\partial r_*}}
\newcommand{\spv}{\frac{\partial}{\partial v_*}}
\newcommand{\spta}{\frac{\partial}{\partial \theta_*}}

\newcommand{\spdr}{\frac{\partial^2}{\partial r_*^2}}
\newcommand{\spdvr}{\frac{\partial^2}{\partial r_* \partial v_*}}
\newcommand{\spdra}{\frac{\partial^2}{\partial r_* \partial \theta_*}}

\begin{document}

\pagestyle{myheadings}
\markboth{}{Wu and Cai}
\draft

\title{\bf Quantum Thermal Effect of Dirac Particles in a
Non-uniformly Rectilinearly Accelerating Black Hole with
Electronic Charge, Magnetic Charge and Cosmological Constant}
\author{S. Q. Wu\thanks{E-mail: sqwu@iopp.ccnu.edu.cn} and
X. Cai\thanks{E-mail: xcai@ccnu.edu.cn}}
\address{Institute of Particle Physics, Hua-Zhong \\
Normal University, Wuhan 430079, China}
\maketitle

\vskip 1cm
\begin{center}{\bf ABSTRACT}\end{center}
\begin{quote}
\widetext
The Hawking radiation of Dirac particles in an arbitrarily rectilinearly
accelerating Kinnersley black hole with electro-magnetic charge and
cosmological constant is investigated by using method of the generalized
tortoise coordinate transformation. Both the location and the temperature
of the event horizon depend on the time and the polar angle. The Hawking
thermal radiation spectrum of Dirac particles is also derived.

\end{quote}

\newpage
\baselineskip 20pt

\section{Introduction}

An important subject of black hole physics is to reveal the thermal properties
of various black holes \cite{Hawk}. Last decades has witnessed much progress
on investigating the thermal properties of scalar fields or Dirac particles in
the stationary axisymmetry black holes \cite{DR,Xuetc}. In the studying of the
Hawking evaporation of the non-stationary black holes, a method of the
generalized tortoise coordinate transformation (GTCT) suggested by Zhao and
Dai \cite{ZD} has been applied to investigate the Hawking thermal radiation
of scalar particles in some non-uniformly accelerating black holes \cite{LZZZ}
and in the non-uniformly accelerating Kerr black hole \cite{WZSZZ}.

However, it is very difficult to investigate the quantum thermal effect of
Dirac particles in the non-stationary black hole. The difficulty lies in
the non-separability of the Chandrasekhar-Dirac equation \cite{CP} in the
most general space-times. The Hawking radiation of Dirac particles in some
non-static black holes has so far been studied in \cite{LZWMY}.

In this paper, we deal with the Hawking effect of Dirac particles in a
non-spherically symmetric and non-stationary Kinnersley black hole with
electronic charge, magnetic charge and cosmological constant \cite{KWT}.
By making use of the GTCT method, we obtain the equation which determines the
event horizon of the Kinnersley black hole. The event horizon equation derived
by the limiting form of Dirac equation near the event horizon is exactly the
same as those given by the null hypersurface which is not spherically symmetric
\cite{LZZZ}. Then we turn to the second order form of the Dirac equation. With
the aid of a GTCT, we adjust the temperature parameter in order that each
component of Dirac spinors satisfies a simple wave equation after being taken
limits approaching the event horizon.

We show that both the shape and the Hawking temperature of the event
horizon of Kinnersley black hole depend on not only the time, but also on
the angle. The location and the temperature coincide with those obtained
by investigating the Hawking effect of Klein-Gordon particles in the
accelerating Kinnersley black hole \cite{LZZZ}.

\section{Dirac equation}

The metric of a non-uniformly rectilinearly accelerating Kinnersley black
hole with electric charge $Q$, magnetic charge $P$ and cosmological constant
$\Lambda$ is given in the advanced Eddington-Finkelstein coordinate system
by \cite{KWT}
\begin{equation}
ds^2 = 2dv(G dv -dr -r^2f d\theta) -r^2(d\theta^2 +\sin^2\theta d\varphi^2)
\end{equation}
where $2G = 1 -\frac{2M}{r} +\frac{Q^2 +P^2}{r^2} -4a\cta\frac{Q^2 +P^2}{r}
-2a r\cos\theta -r^2f^2 -\frac{\Lambda}{3}r^4$, $f = -a\sta$. In the above,
the parameter $a = a(v)$ is the magnitude of acceleration, the mass $M(v)$
and the charges $Q(v), P(v)$ of the hole are functions of time $v$.

We choose a complex null-tetrad $\{\bl, \bn, \bm, \overline{\bm}\}$ such
that $\bl \cdot \bn = -\bm \cdot \overline{\bm} = 1$. Thus the covariant
one-forms can be written as
\begin{equation}
\begin{array}{ll}
\bl &= dv \, ,~~~~ \bn = G dv -dr -r^2f d\theta  \, ,\\
\bm &= \frac{-r}{\sqd}\left(d\theta +i\sta d\varphi\right) \, ,  ~~
\overline{\bm} = \frac{-r}{\sqd}\left(d\theta -i\sta d\varphi\right) \, .
\end{array}
\end{equation}
and their corresponding directional derivatives are
\begin{equation}
\begin{array}{ll}
D &= -\pr \, , ~~~~~~\Delta = \pv +G\pr \, , \\
\delta &= \frac{1}{\sqd r}\left(-r^2f \pr +\pta
+\frac{i}{\sta}\pvi\right) \, , \\
\overline{\delta} &= \frac{1}{\sqd r}\left(-r^2f \pr
+\pta -\frac{i}{\sta}\pvi\right) \, .
\end{array}
\end{equation}

Inserting for the following relations among the Newman-Penrose \cite{NP}
spin-coefficients \footnote{Here and hereafter, we denote $G_{,r} = dG/dr$
, etc.}
\begin{equation}
\begin{array}{lll}
&\epsilon -\rho = -\frac{1}{r} \, ,
&\tilde{\pi} -\alpha = \frac{\coa}{2\sqd r} -\sqd f \, , \\
&\mu -\gamma = \frac{G}{r} +\frac{G_{,r}}{2} \, ,
&\beta -\tau = \frac{\coa}{2\sqd r} -\frac{f}{\sqd} \, ,
\end{array}
\end{equation}
into the spinor form of the coupled Chandrasekhar-Dirac equation \cite{CP},
which describes the dynamic behavior of spin-$1/2$ particles, namely
\begin{equation}
\begin{array}{ll}
&(D +\epsilon -\rho)F_1 +(\overline{\delta} +\tilde{\pi} -\alpha)F_2
= \frac{i\mu_0}{\sqd}G_1 \, , \\
&(\Delta +\mu -\gamma)F_2 +(\delta +\beta -\tau)F_1
= \frac{i\mu_0}{\sqd}G_2 \, ,\\
&(D +\epsilon^* -\rho^*)G_2 -(\delta +\tilde{\pi}^* -\alpha^*)G_1
= \frac{i\mu_0}{\sqd}F_2 \, , \\
&(\Delta +\mu^* -\gamma^*)G_1 -(\overline{\delta} +\beta^* -\tau^*)G_2
= \frac{i\mu_0}{\sqd}F_1 \, ,
\end{array}
\end{equation}
where $\mu_0$ is the mass of Dirac particles, one obtains
\begin{equation}
\begin{array}{rr}
-{\cD}_1 F_1 +\frac{1}{\sqd r}\left({\cL} -r^2f{\cD}_2\right) F_2
&= \frac{i\mu_0}{\sqd} G_1 \, , \\
\left(\pv +G{\cD}_1 +G_{,r}/2\right) F_2
+\frac{1}{\sqd r}\left({\cLd} -r^2f{\cD}_1\right) F_1
&= \frac{i\mu_0}{\sqd} G_2 \, , \\
-{\cD}_1G_2 -\frac{1}{\sqd r}\left({\cLd} -r^2f{\cD}_2\right) G_1
&= \frac{i\mu_0}{\sqd} F_2 \, , \\
\left(\pv +G{\cD}_1 +G_{,r}/2\right) G_1
-\frac{1}{\sqd r}\left({\cL} -r^2f{\cD}_1\right) G_2
&= \frac{i\mu_0}{\sqd} F_1 \, , \label{DCP}
\end{array}
\end{equation}
in which we have defined operators
$${\cD}_n = \pr +\frac{n}{r} \, ,
~~{\cL} = \pta +\frac{1}{2}\coa -\frac{i}{\sta}\pvi  \, ,
~~{\cLd} = \pta +\frac{1}{2}\coa +\frac{i}{\sta}\pvi  \, .$$

By substituting
$$F_1 = \frac{1}{\sqd r} P_1 \, , ~~~F_2 = P_2 \, , ~~~G_1 = Q_1 \, ,
~~~G_2 = \frac{1}{\sqd r} Q_2 \, , $$
into Eq. (\ref{DCP}), they have the form
\begin{equation}
\begin{array}{rr}
-{\cD}_0 P_1 +\left({\cL} -r^2f{\cD}_2\right) P_2
= i\mu_0 r Q_1 \, , &\\
r^2\left(2\pv +2G{\cD}_1 +G_{,r}\right) P_2
+\left({\cLd} -r^2f{\cD}_0\right) P_1
= i\mu_0 r Q_2 \, , &\\
-{\cD}_0 Q_2 -\left({\cLd} -r^2f{\cD}_2\right) Q_1
= i\mu_0 r P_2 \, , &\\
r^2\left(2\pv +2G{\cD}_1 +G_{,r}\right) Q_1
-\left({\cL} -r^2f{\cD}_0\right) Q_2
= i\mu_0 r P_1 \, . &\label{reDP}
\end{array}
\end{equation}

\section{Event horizon}

An apparent fact is that the Chandrasekhar-Dirac equation (\ref{reDP}) could
be satisfied by identifying $Q_1$, $Q_2$ with $P_2^*$, $-P_1^*$, respectively.
So one may deal with a pair of components $P_1$, $P_2$ only. Although Eq.
(\ref{reDP}) can not be decoupled, to deal with the problem of Hawking
radiation, one may concern about the behavior of Eq. (\ref{reDP}) near
the horizon only. As the space-time we consider at present has a symmetry
about $\varphi$-axis, we can introduce the generalized tortoise coordinate
transformation \cite{ZD}
\begin{equation}
\begin{array}{ll}
r_* &= r +\frac{1}{2\kappa}\ln[r -r_H(v,\theta)] \, , \\
v_* &= v -v_0 \, , ~~\theta_* = \theta -\theta_0 \, ,\label{trans}
\end{array}
\end{equation}
where $r_H$ is the location of the event horizon, $\kappa$ is an adjustable
parameter and is unchanged under tortoise transformation. Both parameters
$v_0$ and $\theta_0$ are arbitrary constants. From formula (\ref{trans}),
we can deduce some useful relations for the derivatives as follows:
\begin{eqnarray*}
\begin{array}{lll}
&\pr = \left[1 +\frac{1}{\tkr}\right]\spr \, ,\\
&\pv = \spv -\frac{r_{H,v}}{\tkr}\spr \, , \\
&\pta = \spta -\frac{r_{H,\theta}}{\tkr}\spr  \, .
\end{array}
\end{eqnarray*}

Under the transformation (\ref{trans}), Eq. (\ref{reDP}) with regards to
($P_1,P_2$) can be reduced to the following limiting form near the event
horizon \footnote{Throughout the paper, we make a convention that all
coefficients in the front of each derivatives term take values at the
event horizon when a GTCT is made and followed by taking limits approaching
the event horizon.}
\begin{equation}
\begin{array}{ll}
&\spr P_1 +\left(r_{H,\theta} +r_H^2f \right)\spr P_2 = 0 \, , \\
&-\left(r_{H,\theta} +r_H^2f\right)\spr P_1
+2r_H^2\left(G -r_{H,v}\right)\spr P_2 = 0 \, ,  \label{trDPP}
\end{array}
\end{equation}
after being taken limits $r \rightarrow r_H(v_0,\theta_0)$, $v \rightarrow v_0$
and $\theta \rightarrow \theta_0$. A similar form holds for $Q_1, Q_2$.

If the derivatives $\spr P_1$ and $\spr P_2$ in Eq. (\ref{trDPP}) do not be
equal to zero, the existence condition of non-trial solutions for $P_1$ and
$P_2$ is that the determinant of Eq. (\ref{trDPP}) vanishes, which gives
the following equation to determine the location of horizon
\begin{equation}
2G -2r_{H,v} +r_H^2f^2 +2f r_{H,\theta}
+\frac{r_{H,\theta}^2}{r_H^2} = 0 \, . \label{loca}
\end{equation}
The event horizon equation (\ref{loca}) can be inferred from the null
hypersurface condition, $g^{ij}\partial_i F\partial_j F = 0$, and $F(v,r,
\theta) = 0$, namely $r = r(v,\theta)$. The location of the event horizon
is in accord with that obtained in the case of discussing about the thermal
effect of Klein-Gordon particles in the same space-time \cite{LZZZ}. It
follows that $r_H$ depends not only on $v$, but also on $\theta$. So the
location of the event horizon and the shape of the black hole change with
time.

\section{Hawking temperature}

To investigate the Hawking radiation of spin-$1/2$ particles, one may only
deal with the behavior of $P_1, P_2$ components of Dirac equation near the
event horizon because one can set
\begin{equation}
Q_2 = -P_1^*\, , ~~~~~~ Q_1 = P_2^* \, .
\end{equation}
A direct calculation gives the second-order form of Dirac equation for the
two-component spinor ($P_1, P_2$)
\begin{eqnarray}
&&[r^2(2\pv +2G{\cD}_0 +G_{,r}){\cD}_0 +({\cL} -r^2f{\cD}_{-1})
({\cLd} -r^2f{\cD}_0)] P_1 \nonumber \\
&&~~~~~~~+r^2[-(2\pv +2G{\cD}_0 +G_{,r})({\cL} -r^2f{\cD}_2) \nonumber \\
&&+({\cL} -r^2f{\cD}_1)(2\pv +2G{\cD}_1 +G_{,r})]P_2
= \mu_0^2 r^2 P_1 \, ,\label{socd+}
\end{eqnarray}
and
\begin{eqnarray}
&&[r^2{\cD}_1(2\pv +2G{\cD}_1 +G_{,r}) +({\cLd} -r^2f{\cD}_1)({\cL}
-r^2f{\cD}_2)] P_2 \nonumber \\
&&~~~~~~+[{\cD}_{-1}({\cLd} -r^2f{\cD}_0) -({\cLd} -r^2f{\cD}_1){\cD}_0 ] P_1
= \mu_0^2 r^2 P_2 \, . \label{socd-}
\end{eqnarray}

Given the GTCT in Eq. (\ref{trans}) and after some tedious calculations, the
limiting form of Eqs. (\ref{socd+},\ref{socd-}), when $r$ approaches $r_H(v_0,
\theta_0)$, $v$ goes to $v_0$ and $\theta$ goes to $\theta_0$, reads
\begin{equation}
\begin{array}{ll}
&\left[\frac{A}{2\kappa} +2r_H^2(2G -r_{H,v}) +2r_H^4f^2
+2f r_{H,\theta}r_H^2\right] \spdr P_1 +2r_H^2 \spdvr P_1 \\
&-2\left(f r_H^2 +r_{H,\theta}\right) \spdra P_1
+\left[-A +r_H^2G_{,r} +r_H^3f^2 -r_H^2f\coa_0 \right. \\
&\left. -r_H^2f_{,\theta} -(r_Hf +\coa_0)r_{H,\theta}
-r_{H,\theta\theta} \right] \spr P_1 \\
&+2r_H^2\left[r_H^2f_{,v} +G_{,\theta} -\frac{G r_{H,\theta}}{r_H}
-r_H^2f\left(G_{,r} +\frac{r_{H,v}-2G}{r_H}\right)\right] \spr P_2 = 0 \, ,
\label{wone+}
\end{array}
\end{equation}
and
\begin{equation}
\begin{array}{ll}
&\left[\frac{A}{2\kappa} +2r_H^2(2G -r_{H,v}) +2r_H^4f^2
+2f r_{H,\theta}r_H^2\right] \spdr P_2 +2r_H^2 \spdvr P_2 \\
&-2\left(f r_H^2 +r_{H,\theta}\right) \spdra P_2
+\left[-A +3r_H^2G_{,r} +2r_H(2G -r_{H,v}) +5r_H^3f^2 \right. \\
&\left. -r_H^2f_{,\theta} -r_H^2f\coa_0 +(3f r_H -\coa_0)r_{H,\theta}
-r_{H,\theta\theta} \right] \spr P_2 +\frac{r_{H,\theta}}{r_H}
\spr P_1 = 0 \, . \label{wone-}
\end{array}
\end{equation}

With the aid of the event horizon equation (\ref{loca}), we know that the
coefficient $A$ is an infinite limit of $0 \over 0$ type. By use of the
L' H\^{o}spital rule, we get the following result
\begin{eqnarray}
A &=& \lim_{r \rightarrow r_H(v_0,\theta_0)}
\frac{2r^2(G -r_{H,v}) +r^4f^2 +2f r^2r_{H,\theta}
+r_{H,\theta}^2}{r -r_H} \nonumber\\
&=& 2r_H^2G_{,r} +4r_H(G -r_{H,v}) +4r_H^3f^2 +4f r_Hr_{H,\theta} \nonumber \\
&=& 2r_H^2G_{,r} +2r_H^3f^2 -\frac{2r_{H,\theta}^2 }{r_H} \, .
\end{eqnarray}

Now let us select the adjustable parameter $\kappa$ in Eqs. (\ref{wone+},
\ref{wone-}) such that
\begin{eqnarray}
r_H^2 &\equiv& \frac{A}{2\kappa} +2r_H^2(2G -r_{H,v}) +2r_H^4f^2
+2f r_H^2r_{H,\theta} \nonumber\\
&=& \frac{r_H^3G_{,r} +r_H^4f^2 -r_{H,\theta}^2}{\kappa r_H}
+2Gr_H^2 +r_H^4f^2 -r_{H,\theta}^2 \, ,
\end{eqnarray}
which means the temperature of the horizon is
\begin{equation}
\kappa =\frac{r_H^2G_{,r} +r_H^3f^2 -r_{H,\theta}^2/r_H}{ r_H^2(1-2G)
-r_H^4f^2 +r_{H,\theta}^2} \, . \label{temp}
\end{equation}
Such a parameter adjustment can make Eqs. (\ref{wone+},\ref{wone-}) reduce to
\begin{equation}
\begin{array}{ll}
&\spdr P_1 +2\spdvr P_1 -2\left(f +\frac{r_{H,\theta}}{r_H^2}\right)
\spdra P_1 +\left[r_Hf^2 -G_{,r}  \right. \\
&\left. -f_{,\theta} -f\coa_0 -(r_Hf +\coa_0)\frac{r_{H,\theta}}{r_H^2}
+\frac{2r_{H,\theta}^2}{r_H^3} -\frac{r_{H,\theta\theta}}{r_H^2}\right]
\spr P_1  \\
&+2\left[G_{,\theta} +r_H^2f_{,v} -\frac{G r_{H,\theta}}{r_H}
-r_H^2f\left(G_{,r} +\frac{r_{H,v}-2G}{r_H}\right) \right] \spr P_2 = 0 \, ,
\label{wtwo+}
\end{array}
\end{equation}
and
\begin{equation}
\begin{array}{ll}
&\spdr P_2 +2\spdvr P_2 -2\left(f +\frac{r_{H,\theta}}{r_H^2}\right) \spdra
P_2 +\left[5r_Hf^2 +G_{,r} +\frac{4G -2r_{H,v}}{r_H} \right. \\
&\left. -f_{,\theta} -f\coa_0 +\left(3f r_H -\coa_0\right)
\frac{r_{H,\theta}}{r_H^2} +\frac{2r_{H,\theta}^2}{r_H^3}
-\frac{r_{H,\theta\theta}}{r_H^2}\right] \spr P_2
+\frac{r_{H,\theta}}{r_H^3} \spr P_1 = 0 \, .
\label{wtwo-}
\end{array}
\end{equation}

Using Eq. (\ref{trDPP}), Eqs. (\ref{wtwo+},\ref{wtwo-}) can be recast into
the following standard wave equation near the horizon in an united form
\begin{eqnarray}
\spdr \Psi +2\spdvr \Psi -2C_1 \spdra \Psi + 2C_2 \spr \Psi = 0 \, ,
\label{wave}
\end{eqnarray}
where $C_1, C_2$ will all be regarded as finite real constants,
$$C_1 = f +\frac{r_{H,\theta}}{r_H^2 } \, , $$
\begin{eqnarray*}
2C_2 &=& -r_Hf^2 -G_{,r} -f_{,\theta} -f\coa_0
-(r_Hf +\coa_0)\frac{r_{H,\theta}}{r_H^2} +\frac{2r_{H,\theta}^2}{r_H^3}
-\frac{r_{H,\theta\theta}}{r_H^2}\nonumber\\
&&-\frac{r_H^2f +r_{H,\theta}}{(G -r_{H,v})r_H^3}
\left[G r_{H,\theta} -r_HG_{,\theta} -r_H^3f_{,v}
+r_H^2f\left(G_{,r}r_H +r_{H,v} -2G\right)\right]
\end{eqnarray*}
for $\Psi=P_1$, and
\begin{eqnarray*}
2C_2 &=& 3r_Hf^2 -f_{,\theta} -f\coa_0 +G_{,r} +\frac{4G-2r_{H,v}}{r_H}  \\
&&+\left(2f r_H -\coa_0\right)\frac{r_{H,\theta}}{r_H^2}
+\frac{r_{H,\theta}^2}{r_H^3} -\frac{r_{H,\theta\theta}}{r_H^2}
\end{eqnarray*}
for $\Psi=P_2$.

\section{Thermal radiation spectrum}

Now separating variable as follows
$$\Psi = R(r_*)\Theta(\theta_*)e^{-i\omega v_* +im\varphi}$$
and substituting this into equation
(\ref{wave}), one gets
\begin{equation}
\begin{array}{ll}
&\Theta^{\prime} = \lambda \Theta \, ,\\
& R^{\prime\prime} = 2(i\omega -C_0) R^{\prime} \, ,
\end{array}
\end{equation}
where $\lambda$ is a real constant introduced in the separation variables,
$C_0 = C_2 -\lambda C_1$. The solutions are
\begin{equation}
\begin{array}{ll}
&\Theta = e^{\lambda \theta_*} \, ,\\
& R \sim e^{2(i\omega -C_0)r_*} \, ; R_0 \, .
\end{array}
\end{equation}

The ingoing wave and the outgoing wave to Eq. (\ref{wave}) are
\begin{equation}
\begin{array}{ll}
&\Psi_{\rm in} = e^{-i\omega v_* +im\varphi +\lambda \theta_*} \, ,\\
&\Psi_{\rm out} = e^{-i\omega v_* +im\varphi +\lambda \theta_*}
e^{2(i\omega -C_0)r_*} \, ,~~~~~~~ (r > r_H) \, .
\end{array}
\end{equation}

Near the event horizon, we have
$$r_* \sim \frac{1}{2\kappa}\ln (r - r_H) \, .$$
Clearly, the outgoing wave $\Psi_{\rm out}(r > r_H)$
is not analytic at the event horizon $r=r_H$, but can be analytically
extended from the outside of the hole into the inside of the hole through
the lower complex $r$-plane
$$ (r -r_H) \rightarrow (r_H -r)e^{-i\pi}$$
to
\begin{equation}
\tilde{\Psi}_{\rm out} = e^{-i\omega v_* +im\varphi
+\lambda \theta_*}e^{2(i\omega -C_0)r_*}
e^{i\pi C_0/\kappa}e^{\pi\omega /\kappa} \, ,~~~~~~(r < r_H) \, .
\end{equation}

So the relative scattering probability of the outgoing wave at
the horizon is easily obtained
\begin{equation}
\left|\frac{{\Psi}_{\rm out}}{\tilde{\Psi}_{\rm out}}\right|^2
= e^{-2\pi\omega/\kappa} \, .
\end{equation}

According to the method suggested by Damour and Ruffini \cite{DR} and
developed by Sannan \cite{San}, the thermal radiation Fermionic spectrum
of Dirac particles from the event horizon of the hole is given by
\begin{equation}
\langle N(\omega) \rangle
= \frac{1}{e^{\omega/T_H } + 1} \, , \label{sptr}
\end{equation}
with the Hawking temperature being
$$ T_H = \frac{\kappa}{2\pi} \, ,$$
whose obvious expression is
\begin{equation}
T_H = \frac{1}{4\pi r_H} \cdot \frac{M r_H -r_H^3a\cta_0
+(2r_Ha\cta_0 -1)(Q^2 +P^2) -\frac{\Lambda}{3}r_H^4
-r_{H,\theta}^2}{M r_H +r_H^3a\cta_0 +(2r_Ha\cta_0 -1/2)(Q^2 +P^2)
-\frac{\Lambda}{6}r_H^4 +\frac{r_{H,\theta}^2}{2}} \, .
\end{equation}
It follows that the temperature depends not only on the time, but also on
the angle $\theta$ because it is determined by the surface gravity $\kappa$,
a function of $v$ and $\theta$. The temperature is in consistent with that
derived from the investigating of the thermal radiation of Klein-Gordon
particles \cite{LZZZ}.

\section{Conclusions}

Equations (\ref{loca}) and (\ref{temp}) give the location and the temperature
of event horizon of the hole, which depend not only on the advanced time $v$
but also on the polar angle $\theta$. Eq. (\ref{sptr}) shows the thermal
radiation spectrum of Dirac particles in an arbitrarily rectilinearly
accelerating Kinnersley black hole.

In conclusion, we have studied the Hawking radiation of Dirac particles
in an arbitrarily accelerating Kinnersley black hole whose mass and charges
change with time. The Chandrasekhar-Dirac equation can not be decoupled in
the most general black hole background, however, under the generalized
tortoise coordinate transformation, the limiting form of its corresponding
second order equation takes the standard form of wave equation near the
event horizon, to which separation of variables is possible.

Both the location and the temperature of the event horizon of the accelerating
Kinnersley black hole depend on the time and the angle. They are just the
same as that obtained in the discussing on thermal radiation of Klein-Gordon
particles in the same space-time.

\vskip 0.5cm
\noindent
{\bf Acknowledgment}

S.Q. Wu is very grateful to Dr. Jeff at Motomola Company for his long-term
helps. This work is supported in part by the NSFC in China.


\noindent
\section*{\bf Appendix: Newman-Penrose coefficients}
\vskip 0.3cm
\setcounter{equation}{0}
\renewcommand{\theequation}{A.\arabic{equation}}

The complex null-tetrad $\{\bl, \bn, \bm, \overline{\bm}\}$ that satisfies
the orthogonal conditions $\bl \cdot \bn = -\bm \cdot \overline{\bm} = 1$
in the Kinnersley black hole is chosen as
\begin{equation}
\begin{array}{ll}
\bl &= dv \, ,~~~~ \bn = G dv -dr -r^2f d\theta \, ,\\
\bm &= \frac{-r}{\sqd}\left(d\theta +i\sta d\varphi\right) \, , ~~
\overline{\bm} = \frac{-r}{\sqd}\left(d\theta -i\sta d\varphi\right) \, .
\end{array}
\end{equation}

It is not difficult to determine the twelve Newman-Penrose complex
coefficients \cite{NP} in the above null-tetrad as follows
\begin{eqnarray}
\begin{array}{lll}
&&\tilde{\kappa} = \tilde{\lambda} = \sigma = \epsilon=0 \, ,
~~\rho = \frac{1}{r} \, , ~~\mu = \frac{G}{r} \, , ~~\gamma = -G_{,r}/2 \, ,
~~\tau = -\tilde{\pi} = \frac{f}{\sqd} \, , \nonumber \\
&&\alpha = -\frac{\coa}{2\sqd r} +\frac{f}{\sqd}\, ,
~~\beta = \frac{\coa}{2\sqd r} \, , ~~\nu = \frac{1}{\sqd r}
\left[(2rG -r^2G_{,r})f +r^2f_{,v} +G_{,\theta} \right] \, .
\end{array}
\end{eqnarray}

\end{document}